\documentclass[preprint]{revtex4}

\newcommand{\be}{\begin{equation}}
\newcommand{\ee}{\end{equation}}
\newcommand{\bea}{\begin{eqnarray}}
\newcommand{\eea}{\end{eqnarray}}

\begin{document}
\titlepage

\title{Horizon Entropy in Modified Gravity}

\author{Peng Wang \footnote{pengwang@stanford.edu}}
\affiliation{\it Department of Physics, Stanford University,
Stanford, CA 94305-4060, USA \footnote{This address will be valid
after Sep. 21, 2005}}
\begin{abstract}
We present an observation about the proposal that four-dimensional modification of general
relativity may explain the observed cosmic acceleration today. Assuming that the
thermodynamical nature of gravity theory continues to hold in modified gravity theories, we
derive the modified horizon entropy formula from the modified Friedmann equation. We argue
that our results imply that there are conceptual problems in some models of
four-dimensional
modification of general relativity.
\end{abstract}

\maketitle

It is a well-known result that in general relativity, we can assign an entropy to black
hole horizon \cite{hawking} and cosmological horizon \cite{hawking2} by the same formula
(see e.g. Ref.~\cite{pad} for extended reviews and discussions):
\begin{equation}
S={A\over 4G}\ ,\label{}
\end{equation}
where $A$ is the proper area of the horizon.

On the other hand, Jacobson \cite{jacobson} made an interesting
observation that Einstein equation can be derived by assuming the
universality of formula (1) on any local Rindler horizons.

From those two results, we have the observation that
\emph{four-dimensional gravitational theory and gravitational
entropy formula are very likely in one-to-one correspondence}.

Recently, there are intense discussions on whether the observed
cosmic acceleration is due to the fact that general relativity
will be modified at large scale so that currently the Friedmann
equation is modified \cite{lue, 1/R, dvali, freese, dt}. There are
several different proposals to modify general relativity in the
literature, e.g. extra dimensions \cite{dvali}, higher derivative
curvature terms \cite{1/R}, etc. In this work, we will consider
the possibility that modified Friedmann equation of the form
(\ref{mf}) is the result of four-dimensional modification of
general relativity in the large scale \cite{lue}. In this
framework, the Universe is always matter dominated, but due to
large scale modification of general relativity, matter can cause
the Universe to accelerate today. Then based on the above
observation, it is conceivable that the formula for gravitational
entropy will also be modified in large scale. We will show that
this is indeed the case. Thus, the main result of this paper is
that, when we try to build models of four-dimensional modification
of general relativity to explain cosmic acceleration, we should
bear in mind that the modified gravity theory should reproduce the
modified horizon entropy formula derived in this paper.
Interestingly, some of the modified entropy formula is so
obviously unphysical (e.g. the $\alpha<0$ Cardassian model) that
it tells us that we'd better not spend our time on such endeavor.

As the cosmology driven by modified Friedmann equation is
generally not de Sitter, we should first address the question of
whether can we assign a gravitational entropy to it? If yes, then
since the event horizon, apparent horizon and particle horizon
are different, which one should we consider?

We think the proper choice is the \emph{apparent horizon} which is
the boundary surface of anti-trapped region. In
Ref.~\cite{fischler}, the particle horizon is taken as the
holographic boundary. However, in Ref.~\cite{kaloper}, it is shown
that this choice will violate the holographic bound in inflation.
Indeed, let's assume that inflation expands only $10^{30}$, it
occurs at the GUT scale $H\sim10^{-6}$ (in Planck units) and the
temperature after reheating is $T\sim10^{-3}$. In this case the
size of particle horizon after inflation will be $L_{PH}\sim
H^{-1}\times10^{30}\sim10^{36}$, the area $A\sim
L^2_{PH}\sim10^{72}$ and the entropy $S\sim
T^3L^3_{PH}\sim10^{99}$, which clearly violates the bound $S/A<1$.
Instead, if we consider the holographic bound as the apparent
horizon $L_{A}\sim H^{-1}\sim10^{6}$. Then the area $A\sim
L^2_{A}\sim 10^{12}$ and the entropy $S\sim T^3L^3_{A}\sim 10^{9}$
which clearly satisfies the holographic bound. Thus, insisting on
the validity of holography during inflation, we should choose the
apparent horizon as the holographic boundary.

Moreover, research in black holes also supports the viewpoint
that we should focus on apparent horizon. General accelerating
cosmological spacetimes are quite similar to dynamical black
holes. For dynamical black holes, we also face the question of
whether can we assign gravitational entropy to them and how can
we define horizon. This problem is analyzed several years ago by
Hayward et al. \cite{hayward}. The conclusion is that it is
sensible to assign a gravitational entropy to the trapping
horizon of dynamical black holes defined as hypersurfaces
foliated by marginal surfaces. Moreover, as argued by Jacobson
\cite{jacobson}, the entire framework of black hole
thermodynamics and in particular the notion of black hole entropy
extends to any causal horizon. In cosmological spacetime, the
corresponding object is just the apparent horizon.

Thus, we will focus on the entropy associated with the cosmic
apparent horizon in this work. Let's assume that in modified
gravity theory, the gravitational entropy is still determined
solely by the area of the apparent horizon. So we have
\begin{equation}
S=f(A/4G), \mbox{when} \ L_A>{1\over H_0} .\label{1}
\end{equation}
where $f$ is an arbitrary function which we will show in the below
that it is determined by the modified Friedmann equation, $A=4\pi
L_A^2$ and $L_A=1/H$ are the area and radius of the apparent
horizon. From the discussion below, this assumption is actually
equivalent to the requirement that the modified Friedmann equation
does not contain \emph{derivatives} of $H$.

Since temperature of the horizon is derived from the requirement
that the wick-rotated metric is smooth on the horizon, it is
independent of the gravity theory that leads to the horizon
geometry (see e.g. Ref.~\cite{townsend}). Thus the expression for
temperature should remain the same in modified gravity theory,
which reads $T=1/(2\pi L_A)$.

The heat flux across the apparent horizon is given by
\cite{hayward}
\begin{equation}
dQ=A(\psi_t dt+\psi_r dr),\label{flux}
\end{equation}
where $\psi$ is the energy-supply vector defined as
\begin{equation}
\psi_a=T^b_a\partial_b L_A+w\partial_a L_A\label{}
\end{equation}
where $w$ is the work density defined as $w=T_{ab}h^{ab}$.
$T_{ab}$ is the projection of the 4d energy-momentum tensor
$T_{\mu\nu}$ of a perfect fluid on the normal direction of the
two-sphere defined as $r=\mbox{const.}$.

Then we can find that in FRW metric, the heat flux is given by
\begin{equation}
dQ=A(\rho+p)dt.\label{dq}
\end{equation}
where we have used the fact that $r=1/(aH)$ on the apparent
horizon.

Assuming the thermodynamical nature of gravity continues to hold in
modified gravity theory \cite{jacobson}, we have the relation
$dQ=TdS$. Then from this and equations (\ref{dq}) and (\ref{1}), we
can find that
\begin{equation}
\dot H=-{4\pi G\over f'(A/4G)}(\rho+p).\label{2}
\end{equation}
Note that for $f(x)=x$, what we did above is exactly the
derivation of Einstein equation from gravitational entropy formula
by Jacobson in the special case of Firedmann-Robertson-Walker
metric. It is also important to notice that we did not use the
first law of thermodynamics in deriving Eq.~(\ref{2}) so we
circumvented the question of defining gravitational energy in
general cosmological spacetime, which is still unresolved.

Now let's consider the general form of modified Friedmann equation
studied in Ref.~\cite{lue},
\begin{equation}
H^2=H_0^2g(x),\label{mf}
\end{equation}
where $g$ is an arbitrary function and $x\equiv\rho/\rho_{c0}$
with $\rho_{c0}$ the current critical density.

An example of the form (\ref{mf}) that has received much
discussion in recent literature is the so called ``Cardassian"
cosmology \cite{freese}:
\begin{equation}
H^2={8\pi G\over3}\rho+B\rho^\alpha\ .\label{ca}
\end{equation}
When $\alpha<2/3$, this can drive an accelerating Universe without
introducing dark energy \cite{freese}.

Of course, modified gravity may not be the unique way to arrive at
Eq.~(\ref{mf}). For example, Reference \cite{freese2} proposed
that Eq.~(\ref{ca}) may be the result of exotic interaction
property of dark matter. In this paper we will focus on the
possibility that the modified Friedmann equation (\ref{mf}) is the
result of large-scale modification of general relativity.

From Eq.~(\ref{mf}) and the continuity equation
$\dot\rho+3H(\rho+p)=0$, we can get
\begin{equation}
\dot H=-4\pi Gg'(x)(\rho+p)\ .\label{2.1}
\end{equation}

Comparing Eq.~(\ref{2.1}) with Eq.~(\ref{2}), we can find that $f$
and $g$ is related by
\begin{equation}
f'(A/4G)={1\over g'(x)}.\label{3}
\end{equation}
Since $x$ can be expressed in terms of $A$ from Eq.~(\ref{mf}), we
can find $f$ in terms of $g$ from Eq.~(\ref{3}).

From Eq.~(\ref{3}), we can immediately get an important
conclusion: modified Friedmann equations which will give $g'(x)<0$
are probably physically inconsistent because that means that the
horizon entropy $S=f(A/4G)$ will decrease with increasing horizon
area. This seems to be in contradiction with the (generalized)
second law of thermodynamics. Especially, this will rule out
Cardassian expansion model with $\alpha<0$. Thus, while fitting
with cosmological data allows $\alpha<0$ \cite{zhu}, the
$\alpha<0$ case (if as a result of four-dimensional modification
of general relativity), is conceptually problematic. Furthermore,
it is also interesting to notice that the $g'(x)<0$ case just
corresponds to ``super-accelerating" Universe, i.e. $\dot H>0$.
Thus, building models of super-accelerating cosmology from
four-dimensional modification of general relativity is
problematic. On the other hand, if it is firmly established from
observation that our Universe is indeed super-accelerating today.
Then based on our analysis, the acceleration of our Universe is
probably due to some other mechanisms such as a real dark energy
component.

As an example, let's consider Cardassian cosmology with
$\alpha>0$. From Eq.~(\ref{ca}) and Eq.~(\ref{3}), we can find
that when the second term in Eq.~(\ref{ca}) begins to dominate,
i.e. the Universe begins to accelerate, we have
\begin{equation}
f'(A/4G)=\frac{1}{1+(C{A\over4G})^{{1\over\alpha}-1}} \,\label{4}
\end{equation}
where
\begin{equation}
C=\alpha^{{\alpha\over1-\alpha}}\left({3B\rho_{c0}^{\alpha-1}\over8\pi
G}\right)^{{1\over1-\alpha}}{H_0^2G\over\pi} \ . \label{c}
\end{equation}

Integrating Eq.~(\ref{4}), we can get the modified entropy
formula. For example, for $\alpha=1/3$, we can integrate
Eq.~(\ref{4}) explicitly to give
\begin{equation}
S={\mbox{arctan}(CA/4G)\over C}\ .\label{s}
\end{equation}
while for $\alpha=1/2$, we have
\begin{equation}
S=\frac{\ln (1+CA/4G)}{C}.\label{}
\end{equation}

Thus, as commented at the beginning, any four-dimensional theory
that intends to explain Cardassian expansion must reproduce the
strange entropy formulas above. However, those entropy formulas are
so strange (the entropy does not scale like any geometric property
of the system) that in our point of view, this actually disfavors
the existence of such a theory.

The horizon entropy formula in some other proposed modified
Friedmann equation can also be derived in similar ways. For
example, let's consider the modified Friedmann equation proposed
by Dvali and Turner \cite{dt}:
\begin{equation}
H^2-{H^\alpha\over r_c^{2-\alpha}}={8\pi G\over3}\rho \
.\label{dt}
\end{equation}

Differentiating Eq.~(\ref{dt}) with respect to time and using the
continuity equation, we can get
\begin{equation}
\dot H=-{4\pi G\over 1-{\alpha
H^{\alpha-2}\over2r_c^{2-\alpha}}}(\rho+p)\label{}
\end{equation}

Comparing this to Eq.~(\ref{2}) and using the fact that
$A=4\pi/H^2$, we have
\begin{equation}
f'(A/4G)=1-{\alpha\over2r_c^{2-\alpha}}\left({A\over4\pi}\right)^{1-\alpha/2}.\label{}
\end{equation}
which gives
\begin{equation}
S={A\over4G}-{\alpha\over(4-\alpha)r_c^{2-\alpha}}\left({\pi\over
G}\right)^{\alpha/2-1}\left({A\over4G}\right)^{2-\alpha/2}.\label{ss}
\end{equation}
For $\alpha=0$, Eq.~(\ref{ss}) reduce to the standard formula.
This is expected as the $\alpha=0$ case is just the standard
Friedmann equation with a cosmological constant. For $\alpha=1$,
$S$ will decrease with cosmic expansion which is unphysical. Thus
this case cannot be derived from a four-dimensional theory (but it
can be derived in models with extra dimensions and branes
\cite{dvali}). For $\alpha=-1$, $S$ will scale like $A^{5/2}$ with
cosmic expansion, which is again very strange based on our current
understanding of entropy in thermal field theory.

In conclusion, assuming thermodynamical relation $dQ=TdS$
continues to hold in four-dimensional modification of general
relativity, we derived the modified horizon entropy formula in a
class of modified Friedmann equation. Due to the strange form of
the modified entropy formula, we argue that this actually poses a
problem for attempts in this direction.

Finally, we should also comment that our analysis do not apply to all the current
four-dimensional modified gravity theories. For example, for the $f(R)$ gravity theory,
since the modified Friedmann equation contains derivatives of $H$, we cannot establish a
relationship with entropy formula of the form (\ref{1}). Thus our analysis cannot be
applied to it (entropy and other thermodynamical quantities in $f(R)$ gravity were
discussed in Ref.~\cite{odintsov2}; furthermore, it is recently shown in Ref.~\cite{cai}
that similar argument can apply in Gauss-Bonnet gravity). Furthermore, in braneworld models
where we can also get modified Friedmann equation of the form (\ref{mf}), our analysis also
cannot apply directly since the heat flux in this context may also contain contributions
from the bulk matter. It is interesting to pursue whether can we make an analogous analysis
in braneworld models.

\section*{Acknowledgement}
I would like to thank Sergei D. Odintsov for helpful comments on
the draft.

\end{document}